\documentclass{article}

\title{Governance of Decentralized Autonomous Organizations: Voter characteristics, engagement and power concentration}

\title{Blockchain Governance:\\ An Empirical Analysis of User Engagement on DAOs}

\author{Brett Falk\thanks{fbrett@cis.upenn.edu}, Tasneem Pathan\thanks{tpathan@seas.upenn.edu}, Andrew Rigas, Gerry Tsoukalas\thanks{gerryt@bu.edu}}

\input{macros}
\newcommand{\CompoundSD}{April 2020\xspace}
\newcommand{\CompoundED}{December 2022\xspace}
\newcommand{\CompoundNP}{142\xspace}

\newcommand{\UniswapSD}{October 2020\xspace}
\newcommand{\UniswapED}{December 2022\xspace}
\newcommand{\UniswapNP}{30\xspace}

\newcommand{\LidoSD}{December 2020\xspace}
\newcommand{\LidoED}{December 2022\xspace}
\newcommand{\LidoNP}{150\xspace}

\newcommand{\AaveSD}{December 2020\xspace}
\newcommand{\AaveED}{December 2022\xspace}
\newcommand{\AaveNP}{131\xspace}

\begin{document}
\pagestyle{plain}

\maketitle

\begin{abstract}
In this note, we examine voting on four major blockchain DAOs: Aave, Compound, Lido and Uniswap.
Using data directly collected from the Ethereum blockchain, we examine voter activity.

We find that in most votes, the ``minimal quorum,'' i.e., the smallest number of active voters who could swing the vote is quite small.

To understand who is actually driving these DAOs, we use data from the Ethereum Name Service (ENS), Sybil.org, and Compound, to divide voters into different categories.
\end{abstract}

\section{Introduction}

The success of decentralized blockchain-based organizations hinges on their ability to break free from traditional centralized structures and establish efficient self-governance mechanisms. Decentralized Autonomous Organizations (DAOs) whereby a central decision-maker is replaced by automated smart contracts which 
aggregate the votes of token-holders.  This structure has the potential to decentralize decision-making, and allow anyone (who holds governance tokens) to participate in the management of the organization.

To date, there have been many successful DAO implementations across a wide range of areas. 
In the gaming industry, blockchain-native games like Axie Infinity have created an ecosystem of players, developers, and investors, and the governance of the platform is in the hands of the Axie DAO.
Philanthropic DAOs crowdsource donations and decentralizes the process of choosing beneficiaries \cite{philanthropicDAO}.
In finance, for instance, Decentralized Exchanges (DEXs), such as Uniswap offer a viable alternative to centralized exchanges, enabling peer-to-peer decentralized trading.
Similarly, lending protocols like Aave and Compound also use a DAO structure to govern the decision-making of the protocol.
Today, most DeFi projects use a DAO structure to decentralized their governance process.

While DAOs have demonstrated their potential to revolutionize traditional organizational structures, they are not without their challenges. Their decentralized nature can often result in increased inefficiencies and lack of accountability when things go wrong. This can lead to delays in decision-making, conflicts of interest, and difficulties in enforcing rules and regulations. Additionally, while there have been successful DAOs in various sectors, there have also been notable counterexamples (discussed further below). Many DAOs have been criticized for not being ``decentralized enough'' and have struggled with issues such as internal governance disputes, community fragmentation, and low adoption rates. 

These challenges highlight the need for further research. In this paper, we seek to shed light on this innovative organizational form, by providing an empirical investigation of four popular DAOs, discussing their key features, and going over their potential applications and challenges.

Leveraging the public nature of the Ethereum blockchain, we collected the entire voting history for four different DeFi DAOs, Uniswap, Compound, Aave and Lido.  We describe each of these platforms in more detail in Section~\ref{sec:case_studies}.

In Section~\ref{sec:empirical}, we examine the concentration of voting power in each of the four DAOs of interest.  We find that in all four DAOs, voting power is highly concentrated.  As a measure of concentraction, we introduce the notion of the ``minimal quorum size,'' (Definition~\ref{def:minimal_quorum}) which we define to be the fewest number of addresses that could unilaterally decide the outcome of the vote.  In many cases the minimal quorum size is around 3, i.e., if the top three voting stakeholders agree, their preference will win out.  On the other hand, if we consider ``robustness'' as measured by the cost required for an outside attacker to purchase governance tokens in an attempt to influence the outcome of the vote, we find the systems are quite robust, i.e., the dollar cost of buying off a vote is extremely high.

In Section~\ref{sec:whois}, we do a deeper dive into the identity of the voters in these systems.  Using data from the Ethereum Name Service (ENS), Sybil.org and Compound's Governance API we were able to assign real-world identities to the on-chain addresses that cast votes.
From there, we categorized these addresses (e.g. VC, University club, Founder, DeFi Developer, etc.)  This allows us to accurately identify who really controls these protocols.
\section{On-chain governance}

The first major attempt at on-chain governance was the ill-fated  Ethereum DAO (``The DAO''), which aimed to be a decentralized, open investment club \cite{slockitwhitepaper}.  Although ``The'' DAO was a spectacular failure \cite{slockitlessons}, the notion of decentralized, on-chain governance has been steadily growing in popularity, and the majority of DeFi protocols implement some form of on-chain governance.  There are two main reasons why DeFi protocols implement on-chain governance, including decentralization and fundraising, which we describe in more detail below.

\subsection{Decentralization to avoid regulation}

Decentralized Finance (DeFi) applications like on-chain lending platforms (e.g. Aave, Compound, Maker) and on-chain exchanges (e.g. Uniswap, Balancer, Curve) are all controlled by on-chain smart contracts.  Although, in principle, the on-chain smart contracts that govern DeFi protocols like Compound and Uniswap could run forever on the Ethereum blockchain without intervention from the developer team or anyone else, in practice, the core team is needed to add new features, adjust parameters to fit changing market conditions, and make decisions about how to invest the protocol's revenue.

This continued reliance on the core team to continuously shepherd the protocol adds a significant point of centralization, and  ``in cryptocurrency circles, calling something centralized is an insult'' \cite{O18}. One of the primary fears of centralization is that centralized entities are vulnerable to regulation. Puja Ohlaver, a lawyer and strategy counsel for Flashbots describes it this way:

\begin{quote}
``the whole, sort of, premise and justification for crypto is decentralization, right, it's what protects it from being regulated by, you know, antitrust and also securities regulation.''\footnote{\url{https://youtu.be/lKKgP2wS39U?t=1123}}
\end{quote}

Kain Warwick, the founder of Synthetix echoes this sentiment:

\begin{quote}
``when you look at what a DAO enables, you have the ability to kind of transcend any one regulatory jurisdiction or like any legal structure''\footnote{\url{https://youtu.be/dZ4o6wma4js?t=440}}.
\end{quote}

The ability to avoid securities regulation is important to platforms like Synthetix which have long drawn the eye of securities regulators \cite{SynthetixSEC}.
Former SEC director Bill Hinman suggests that decentralized governance may actually be sufficient to avoid securities regulation:

\begin{quote}
``If you have organization where there's not an identifiable active participant or sponsor or third party, that's really directing how the DAO success might go, then you might not have something that gets regulated as a security.''\footnote{\url{https://www.rev.com/transcript-editor/shared/oza8OVR4Zn8w2ilySNlvFkXyr8oWsWHw6wAWw8izOucdSaNwmDEQn2Ahb0CG1ut17UrZw8MOnl0mNtOAjmzlR8CKWYw?loadFrom=PastedDeeplink&ts=422.48}}
\end{quote}

When the lending protocol, Compound, introduced on-chain governance they claimed it was to reduce the protocol's dependence on the initial team, which would make the protocol ``indestructible'' \cite{compoundGovernance}, or ``unstoppable'' \cite{compoundGovernance2}  When Uniswap introduced its governance module (which was a fork of Compound's) they claimed it would make their protocol ``self-sustaining'' and (like Compound) Uniswap would become ``indestructible'' \cite{unitoken}.

Although decentralization is often put forward as a motivation for creating a DAO structure, critics argue that this decentralization is a sham.
An analysis of MakerDAO governance concluded: ``decentralization in DeFi platforms is an illusion'' \cite{makerdecentralization}.
Similarly, ``the DeFi protocols Compound and Uniswap is extremely centralized and controlled by a very small number of addresses'' \cite{FMW22}
The fact that these DAOs are often so centralized caused Andre Cronje to call the term ``Decentralized Finance'' disingenuous \cite{cronje}. 

Although some see the centralization of power in DAO governance as a flaw in the system, others have argued that it's not a flaw -- instead onchain governance systems were set up to give the appearance of decentralization while keeping control centralized among the Venture Capitalists.

\begin{quote}
``it's one-token-one-vote and these large entities like a16z have that many tokens because they purchased them from the DAO ... so I don't think it's a situation where the system is broken per se, because it's working as it was intended the problem comes when they pitch these things as decentralized.''\footnote{\url{https://youtu.be/_ORU-lmEj94?t=230}}
\end{quote}

This question of whether early investors control DAOs came to a head in the proposal to deploy Uniswap v3 on the BNB chain.  The venture capital firm, a16z, voted against the proposal, but was ultimately overruled by the community (including a number of its own delegates).
A more detailed description of this controversy is in Appendix~\ref{app:BNB}.

One possible explanation is that when new DAOs are launched, governance tokens are concentrated among the founders and VCs, but over time these tokens will make their way into the hands of the general population and decentralization will increase.
A recent analysis, however, found that in fact the opposite is the case

\begin{quote}
``DeFi's governance is timocratic. DeFi's timocratic rulers -- the above average token-holders -- accumulate more power over time by purchasing even more tokens.'' \cite{BSP22}
\end{quote}

The appearance of decentralization has also not been completely effective in keeping regulators at bay.
One of the most notable examples of regulatory action against a DAO was the case of Ooki DAO.

In 2021, the exchange, bZx created a DAO structure stating: ``We're going to be really preparing for the new regulatory environment by ensuring bZx is future-proof.''  A claim that was later used against them in the complaint filed by the CFTC \cite{ookiComplaint}.  The CFTC lawsuit against the bZx DAO (later rebranded Ooki DAO) claimed that all voters who participated in Ooki DAO governance were liable for the platform's (many) regulatory violations (e.g. offering leveraged financial products without any form of KYC/AML).  This idea that DAO voters could be held liable for the DAO's misdeeds created fear among DAO participants\footnote{\url{https://youtu.be/0wAALQdaxzA?t=247}}, but does not seem to have discouraged individuals from voting on other platforms.

\subsection{Governance tokens as a fundraising tool}

Just like an IPO, creating governance tokens is also an effective fundraising strategy.  In many cases, projects promise investors a fraction of their governance tokens.  For example, Uniswap gave 17.8\% of its supply of UNI tokens to investors \cite{unitoken}, and Lido gave 22.18\% of its tokens to investors \cite{boardroomLido}.  Other projects, use the newly created tokens to incentivize growth.  Compound used its newly created governance token, COMP, to incentivize users to use the protocol, successfully grew the protocol to one of the largest in DeFi \cite{yieldfarming}.

Why are the tokens valuable?  One common explanation is that governance token holders will eventually vote to direct a portion of future revenues from the platform to themselves \cite{AroundTheBlock}.  In any case, creating a governance token is an easy way to raise capital in the short term to fund development of the protocol.

If governance tokens are akin to equity shares, (offering governance rights and dividends), then perhaps they should be regulated as securities.  Many people think so, and Nicholas Weaver, a prominent security researcher and blockchain critic describes governance tokens as a conspiracy:

\begin{quote}
``basically every DAO governance contract sold on a blockchain is an unlicensed security and they're breaking security law so: conspiracy.''\footnote{\url{ https://youtu.be/J9nv0Ol-R5Q?t=3334}}
\end{quote}

An in-depth legal analysis of Uniswap's UNI token concluded that it is likely a security \cite{kim21}, but other lawyers have argued that some governance tokens (but not others) are likely securities \cite{B22}.  In this work, we largely ignore the question of whether governance tokens are securities, and instead examine the details of their use.

\section{Related Work}

There is a rapidly growing literature on DAOs in general.  Much of this work focuses on the societal and legal implications of DAOs.

On the empirical side, several works have analyzed voting behavior in popular DAOs.  In this category, \cite{AG23} compares decentralization to token prices.
\cite{HLL23} categorizes DAOs based on their governance structure and goals.
\cite{FMW24} Analyzes voting power in the Compound, Uniswap and ENS DAOs.
\cite{SSS24} Analyzes voting in MakerDAO.
\cite{FFHVW24} Gives an overview of different attacks on DAOs.

\section{Case studies}
\label{sec:case_studies}

In this work, we examine the on-chain governance of four of the most popular DeFi protocols on the Ethereum blockchain, Compound, Uniswap, Aave and Lido.  The transparent nature of the blockchain allows us to gather information about voter behavior on each of these platforms.

These four platforms, like most onchain DAOs have a multi-step voting process.  When a user wants to create a proposal, they begin with an off-chain ``temperature check'' through a platform like Snapshot\footnote{\url{https://snapshot.org}}.
Snapshot allows users to use their signing keys to digitally sign votes \emph{that are not submitted to the blockchain}.  Digital signatures allow Snapshot to verify the provenance of each vote, but omitting to submit these votes to the blockchain means that Snapshot could 
censor voters during these temperature checks.  If a proposition is perceived to have enough support, then it can be submitted on-chain for a formal vote.  In order to create a formal (on-chain) proposal, users need to have a minimum number of tokens (e.g. 2.5M UNI\footnote{\url{https://docs.uniswap.org/concepts/governance/process\#phase-3-governance-proposal--governance-portal}}, 25,000 COMP\footnote{\url{https://docs.compound.finance/v2/governance/}}, .5\% or 2\% of the circulating supply of 16M AAVE tokens \footnote{\url{https://docs.aave.com/developers/v/2.0/protocol-governance/governance\#proposition_threshold}}).

Voting for the onchain proposal will last for a period of days (7 days on Uniswap, 3 days on Compound, 3 or 10 days on Aave and 7 days on Lido).  Since participation in these systems is low, in order for a vote to pass, at least some minimum fraction of the token supply must vote (4\% of the voting supply on Uniswap and Compound, 2\% and 20\% on Aave and 5\% on Lido). 

\subsection{Aave}

Aave is a ``decentralized non-custodial liquidity protocol\footnote{\url{https://docs.aave.com/hub/}}.'' It enables users to supply and borrow cryptoassets on Ethereum. Unlike previously introduced decentralized lending strategies that were peer-to-peer, in Aave a borrower can pay interest to borrow a cryptoasset from a liquidity pool smart contract that was initially supplied by another protocol user aiming to earn from interest. Aave was also the first protocol to popularize the concept of a flash loan \cite{flashloan}; this is a feature that allows users to borrow any available amount of an asset without putting up collateral, if the same asset amount is returned to the liquidity pool within the same transaction.

Aave is currently the largest DeFi lending protocol with around \$4B worth of assets available for borrowing\footnote{\url{https://defillama.com/protocol/aave-v2}}.

Aave's governance makes decisions about risk policies such as base risk parameters for overcollateralization and liquidation, improvement policies to enhance various aspects of the protocol such as its own governance contracts, and incentives policies to reward participants in the Aave ecosystem\footnote{\url{ https://docs.aave.com/aavenomics/policies}}.

Between \AaveSD and \AaveED, the Aave DAO has conducted \AaveNP votes.

\subsection{Compound}

Like Aave, Compound is a decentralized lending platform, where users can borrow and lend cryptocurrencies without any form of background check.
Compound is currently the second largest lending protocol on Ethereum with around \$2B worth of assets available for borrowing\footnote{\url{https://defillama.com/protocol/compound-finance}}.
Compound was founded in 2018 by Robert Leshner, who maintains an active role in governing Compound, as well as other DeFi protocols we analyzed \cite{robertleshnervoting}.

Compound's initial VC funding came from a16z, Polychain and Bain \cite{CompoundSeed}.  Compound was one of the first DeFi protocols to introduce token-based governance and the Compound governance system has been widely copied \cite{compoundgovernor}.  
In particular, the Uniswap governance contracts are a fork of Compound's governance contracts.

Compound was one of the first DeFi protocols to introduce governance tokens, with the stated goal: 
``To create unstoppable, upgradable financial infrastructure, Governance replaced our team as administrator of the Compound protocol; COMP token-holders and their delegates debate, propose, and vote on all changes to Compound.'' \cite{compoundGovernance2}

Between \CompoundSD and \CompoundED, the Compound DAO has conducted \CompoundNP votes.

\subsection{Lido}

Lido is a liquid staking service.  Proof of Stake blockchains (like Ethereum) require users to ``stake'' tokens in order to participate in the consensus process.  Users can deposit Ether (or other staking assets like SOL or DOT) with Lido, and Lido will delegate those assets to Proof-of-Stake validators, and reshare the staking rewards with the original depositors.

On Ethereum, Lido users deposit ETH, and receive a staking derivative, stETH, which can be traded freely while the underlying ETH are locked in Ethereum's staking contract.  Lido's ``liquid staking'' allows users to earn staking yield, while holding liquid assets in the form of stETH tokens \cite{lidowhitepaper}.  After Ethereum's ``Shanghai'' upgrade, stETH holders will be able to redeem stETH for the underlying ETH tokens \cite{shanghai}.

Lido DAO manages ``protocol parameters, lists of node operators and oracle members, and can vote on app upgrades.'' \cite{lidomembermanual}  Unpacking this further, protocol parameters include protocol fees (currently at 10\% \cite{lidofees}) and how the fees are split between the Lido Treasury and the node operators.  Node operators means choosing who is actually running the Lido Node infrastructure.  Lido maintains a current list of node operators\footnote{\url{https://fees-monitoring.lido.fi/}}.  Before the merge, Lido node operators were validating on the Ethereum beacon chain, whereas the core Lido contracts were on the PoW Ethereum blockchain.  Oracles were required to pass information about from the beacon chain back to the Ethereum PoW chain\footnote{\url{https://docs.lido.fi/contracts/lido-oracle/}}.  Finally, app upgrades means voting on upgrades to the Lido system, e.g. replacing the Lido oracles with a trustless ZK proof system \footnote{\url{https://research.lido.fi/t/zk-proof-total-value-locked-oracle/3726}}.

Between \LidoSD and \LidoED, the Lido DAO has conducted \LidoNP votes.  Of the four platforms we analyzed, Lido's voters are the least identifiable. 

\subsection{Uniswap}

Uniswap is one of the largest Decentralized Exchanges (DEXs), with trading volumes consistently above \$1B per day\footnote{\url{ https://defillama.com/dex/uniswap}}.
Uniswap's governance began as a fork of Compound's governance platform\footnote{\url{https://github.com/Uniswap/governance}}, and the two remain very similar.

Uniswap's governance structure has been criticized for being overly centralized \cite{Dilendorf}.

A common governance proposal is whether to deploy Uniswap on other blockchains (e.g. Moonbeam\footnote{\url{https://www.tally.xyz/governance/eip155:1:0x408ED6354d4973f66138C91495F2f2FCbd8724C3/proposal/19}}, Gnosis\footnote{\url{https://www.tally.xyz/governance/eip155:1:0x408ED6354d4973f66138C91495F2f2FCbd8724C3/proposal/20}} or Celo\footnote{\url{https://www.tally.xyz/governance/eip155:1:0x408ED6354d4973f66138C91495F2f2FCbd8724C3/proposal/16}}).  Another common type of proposal is adding new fee tiers, i.e., to give liquidity providers the option to charge different fees\footnote{\url{https://www.tally.xyz/governance/eip155:1:0x408ED6354d4973f66138C91495F2f2FCbd8724C3/proposal/14}}.

Between \UniswapSD and \UniswapED, the Uniswap DAO has conducted \UniswapNP votes.

\begin{table}
\caption{Overview of voting activity}
\label{tab:summary}
\begin{tabular}{|l|c|c|c|}
\hline
Platform & Number of Proposals & Avg Number of Voters & Avg Pct of 
 Total Token Supply Cast \\ \hline

Aave & 130 & 121.8 & 3.2\% \\ \hline
Compound & 141 & 72.0 & 7.7\% \\ \hline
Lido & 149 & 6.5 & 5.6\% \\ \hline
Uniswap & 29 & 773.3 & 4.6\% \\ \hline

\end{tabular}
\end{table}

\section{Empirical findings}
\label{sec:empirical}

\subsection{Engagement}

All four platforms we analyzed have low engagement.
There are two natural methods for analyzing engagement, what fraction of the total supply of \emph{tokens} participates in a vote, 
and what fraction of the total number of possible \emph{voters} participates.  Since the votes are all token-weighted, 
we measure engagement by considering what fraction of the tokens voted on any give proposal.

In Table~\ref{tab:summary}, we see that the engagement for all four platforms is low, with only about 5\% of the tokens 
being cast in on any given proposal.

\subsection{Centralization}

There are many ways to characterize ``influential'' voters, but we focus on the notion of a ``minimal quorum'' (Definition~\ref{def:minimal_quorum}).
For a given proposal, we define the ``minimal quorum'' to be the smallest subset of voters who cast votes in that proposal whose combine vote weights are enough to carry the proposal.

\begin{definition}[Minimal Quorum]
\label{def:minimal_quorum}
    Suppose voters $v_1,\ldots,v_n$, with token weights $w_1,\ldots,w_n$ vote on a proposition, and assume $w_1 \ge w_2 \ge \cdots \ge w_n$.  
    The \emph{minimal quorum} is defined to be $v_1,\ldots,v_t$ where $t$ is the smallest integer such that $\sum_{i=1}^t w_i \ge \frac{1}{2} \sum_{i=1}^n w_i$.
\end{definition}

\begin{figure}
    \centering
    \includegraphics[width=.8\textwidth]{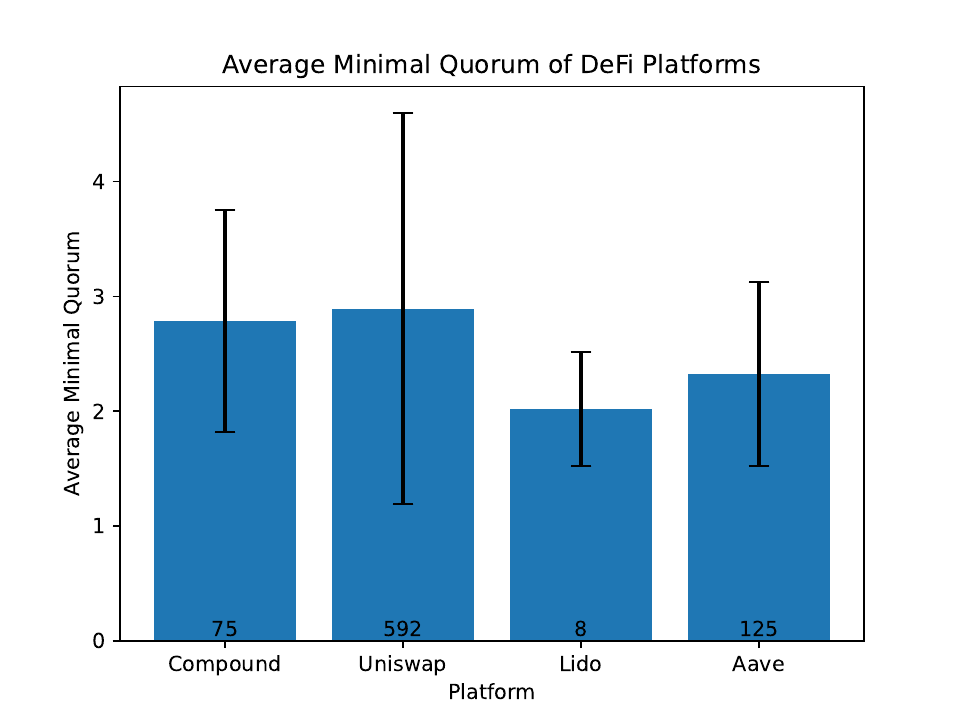}
    \caption{In all the platforms we analyzed, the top 3 or 4 token holders could unilaterally decide almost every vote.  The numbers inside the bars represent the average number of voters who voted on a given proposal.}
    \label{fig:min_quorum}
\end{figure}

Figure~\ref{fig:min_quorum} shows that in most votes, the top three or four token holders together constitute a majority of the voting population.

\subsection{Robustness}

When a DeFi project opens up its governance to the community, there is always a risk of a hostile takeover.
A competing project, or a malicious actor could try to subvert the governance process.
How \emph{robust} are these DAOs to this type of attack?

One way to measure the robustness of a DAO is to measure the cost of purchasing enough governance tokens to change the outcome of the vote 
(assuming that existing voters did not change their votes).
For this analysis, we gathered the historical USD cost of all four governance from the Coingecko API\footnote{\url{https://www.coingecko.com/en/api}}, 
and calculated the cost of buying enough tokens to sway the vote for each proposal on the four platforms we analyzed.

\begin{figure}
    \centering
    \includegraphics{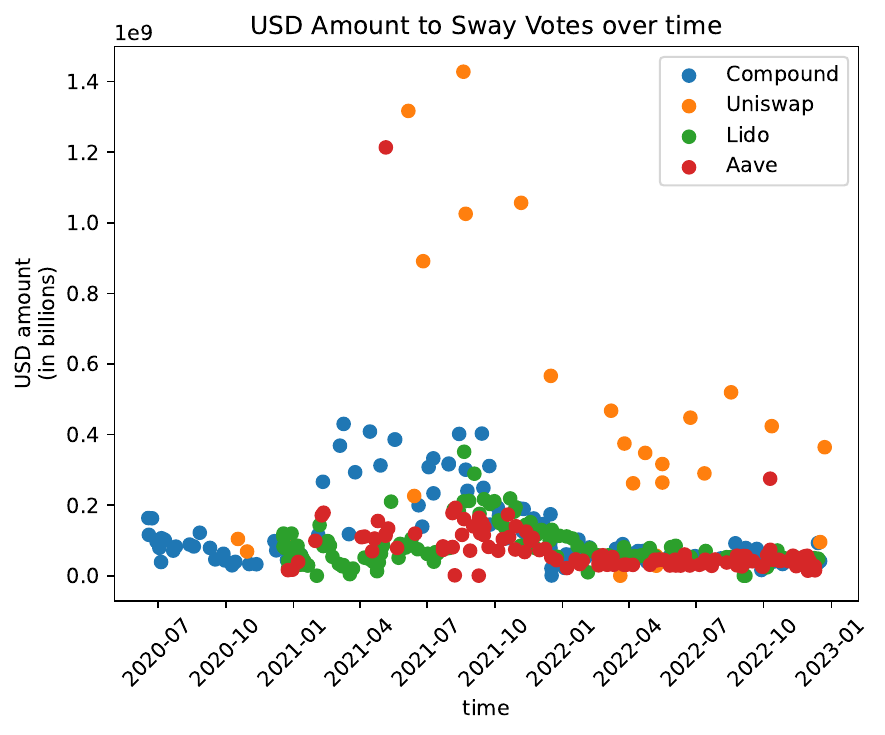}
    \caption{The amount of tokens (denominated in USD) an attacker would need to purchase to sway a vote, assuming the existing voters did not change their position.  The ``robustness'' of the protocols does not seem to be increasing over time.}
    \label{fig:usd_sway}
\end{figure}

In Figure~\ref{fig:usd_sway}, we measure the cost (in USD) of purchasing enough governance tokens to sway the vote on any given proposal.
These numbers are quite high -- it would cost hundreds of millions of dollars to sway the votes on most of these proposals.

Even though the voter turnout is extremely low, the market cap of these governance tokens is very high (As of

\subsection{Contientiousness}

\begin{figure}[H]
    \centering
    \begin{subfigure}[b]{0.49\textwidth}
        \centering
        \includegraphics[width=250pt]{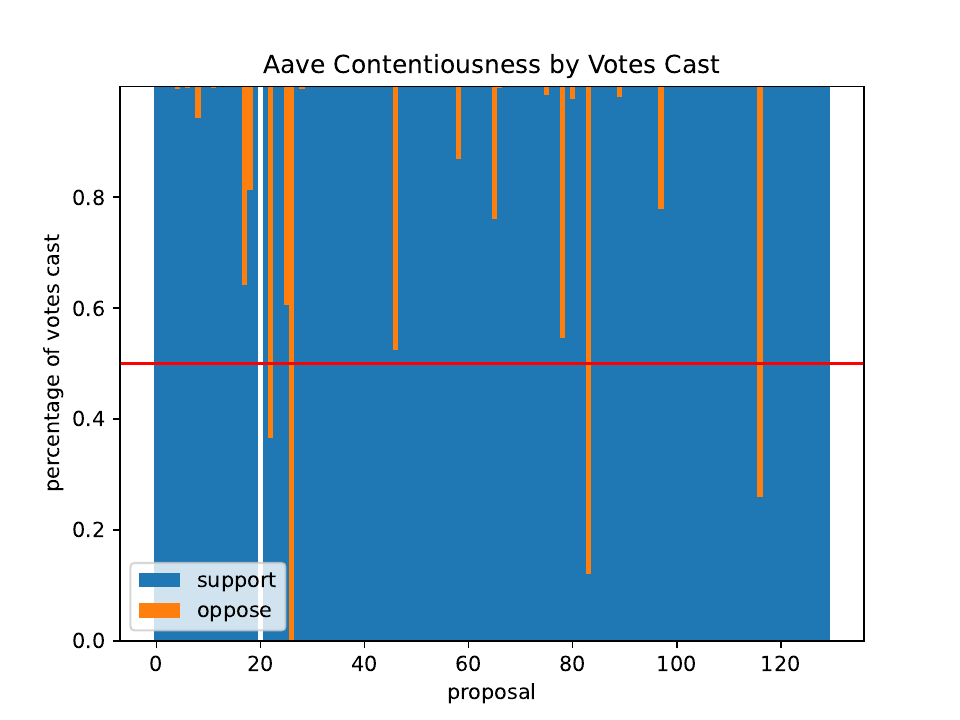}
        \caption{Contientiousness in Aave votes \label{fig:aave_contentiousness}}
    \end{subfigure}
    \hfill
    \begin{subfigure}[b]{0.49\textwidth}
        \centering
        \includegraphics[width=250pt]{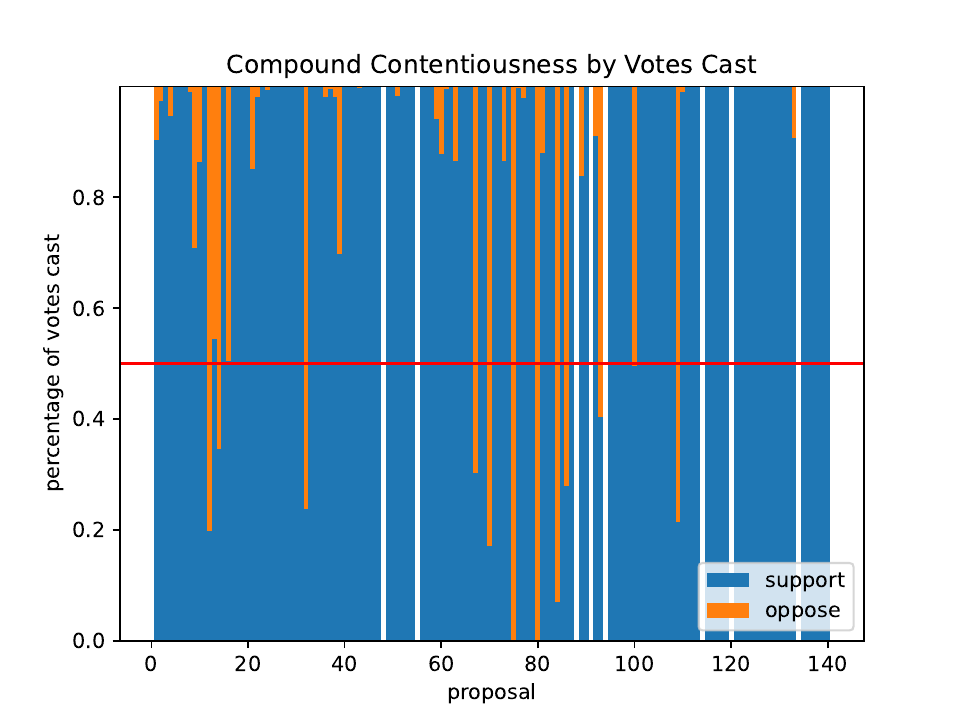}
        \caption{Contientiousness in Compound votes \label{fig:comp_contentiousness}}
    \end{subfigure}
    \begin{subfigure}[b]{0.49\textwidth}
        \centering
        \includegraphics[width=250pt]{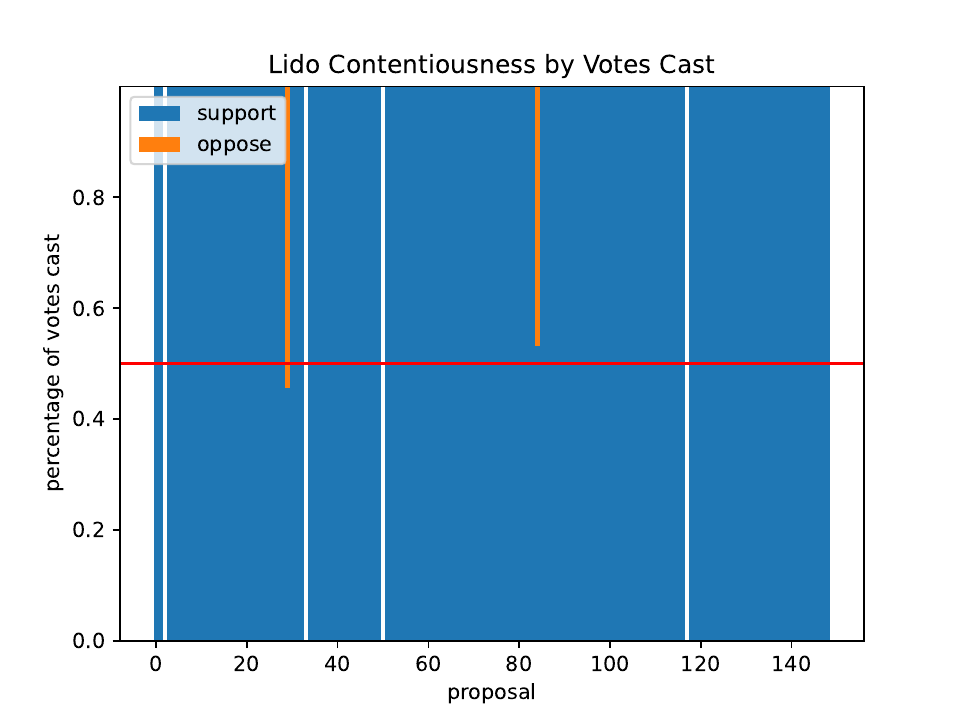}
        \caption{Contientiousness in Lido votes \label{fig:lido_contentiousness}}
    \end{subfigure}
    \hfill
    \begin{subfigure}[b]{0.49\textwidth}
        \centering
        \includegraphics[width=250pt]{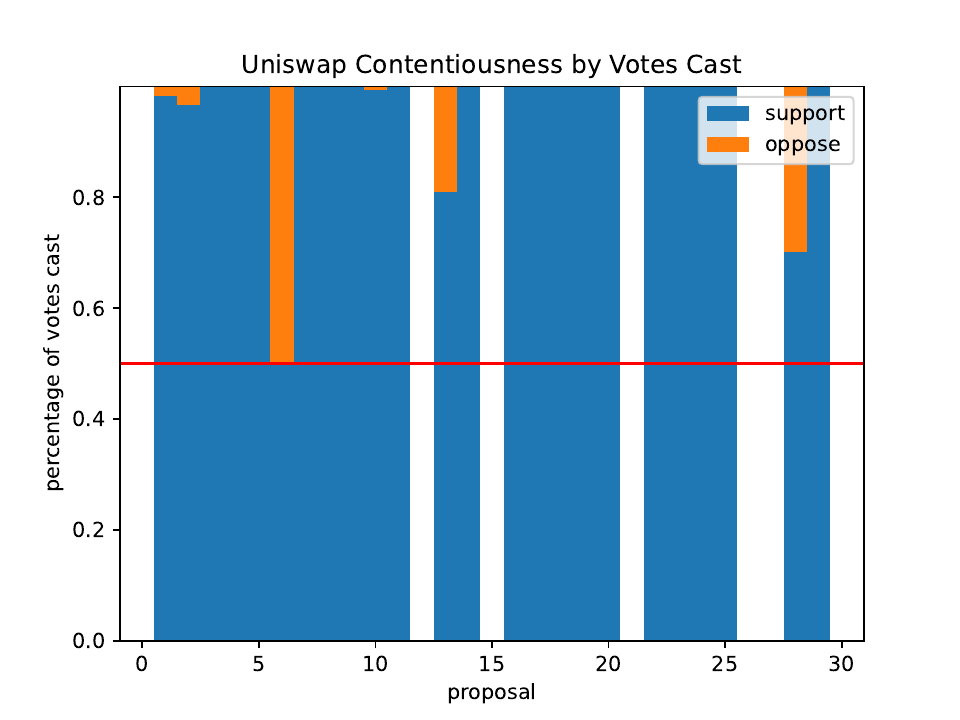}
        \caption{Contientiousness in Uniswap votes \label{fig:uni_contentiousness}}
    \end{subfigure}
    \caption{Because of the off-chain ``temperature checks'' that precede the on-chain votes, almost all votes pass by a large margin.}
\end{figure}

\section{Who is voting?}
\label{sec:whois}

The public nature of the Ethereum blockchain makes it possible to view the Ethereum address of every voter, but these addresses do not, in themselves, carry any information about the off-chain entities that control the account.

In order to associate keys with off-chain actors, we used three primary sources, the Ethereum Name Service (ENS)\footnote{\url{https://docs.ens.domains/}}, Sybil.org's list of verified mappings from Ethereum addresses\footnote{\url{https://github.com/Uniswap/sybil-list}} to social-media profiles (mainly Twitter), as well as Compound's API\footnote{\url{https://docs.compound.finance/v2/api/}}.

Using a combination of these three data sources, we were able to label  the majority of the most powerful voters on Compound and Uniswap.  Unfortunately, many voters on Aave and the majority of voters on Lido remain anonymous.  Some of the largest voters include venture capital funds like a16z and Pantera, University Blockchain Clubs, like those at Berkeley, Columbia, UCLA, Michigan and Penn, as well as individuals like Robert Leshner, the founder of Compound.  See Table~\ref{tab:category_counts}.

\begin{table}
\centering
\caption{Number of proposals for which the voter was a member of the minimal quorum}
\label{tab:category_counts}
\begin{tabular}{|l|c|c|c|c|}
\hline
Voter & Compound & Uniswap & Lido & Aave \\ \hline

Gauntlet & 66 & 10 & 0 & 1 \\ \hline
Polychain Capital & 61 & 0 & 0 & 0 \\ \hline
a16z & 44 & 15 & 0 & 0 \\ \hline
Robert Leshner & 53 & 4 & 0 & 0 \\ \hline
University Clubs & 5 & 15 & 0 & 30 \\ \hline
blck & 46 & 0 & 0 & 0 \\ \hline
MonetSupply & 20 & 0 & 0 & 0 \\ \hline
Uniswap Grants Program & 0 & 7 & 0 & 0 \\ \hline
Bain Capital Ventures & 5 & 0 & 0 & 0 \\ \hline
Kain Warwick & 4 & 0 & 0 & 0 \\ \hline
Pantera Capital & 0 & 0 & 0 & 3 \\ \hline

\end{tabular}
\end{table}

\begin{figure}[H]
    \centering
    \begin{subfigure}[b]{0.45\textwidth}
        \centering
        \includegraphics[width=250pt]{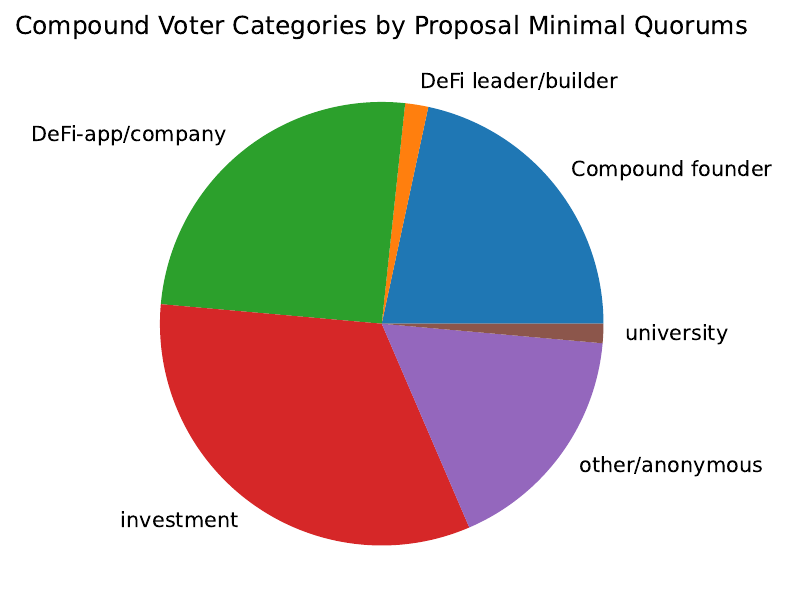}
        \caption{Participation in minimal quorums on Compound, as averaged across all \CompoundNP proposals.  \label{fig:comp_quorum_pie}}
    \end{subfigure}
    \hfill
    \begin{subfigure}[b]{0.45\textwidth}
        \centering
        \includegraphics[width=250pt]{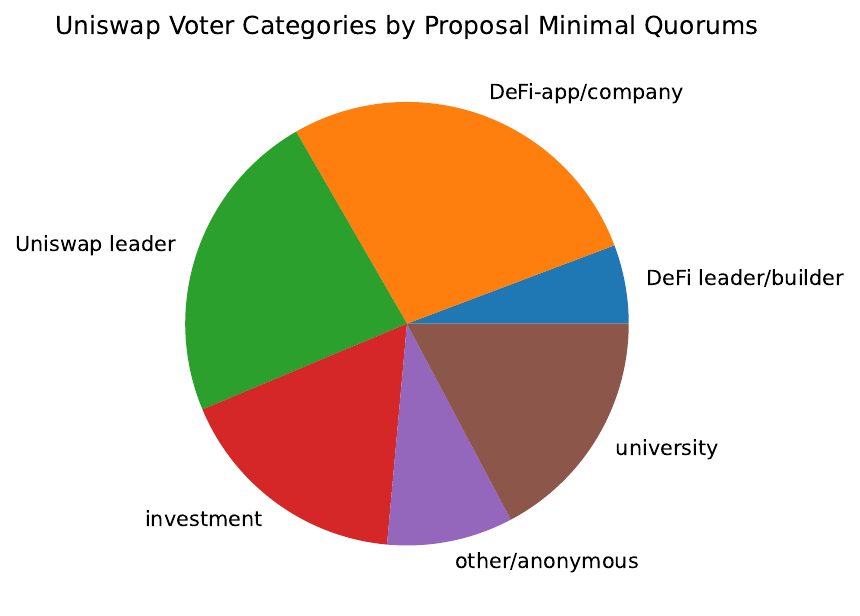}
        \caption{Participation in minimal quorums on Uniswap, as averaged across all \UniswapNP proposals.\label{fig:uniswap_quorum_pie}}
    \end{subfigure}
    \caption{Investment firms (venture capitalists like a16z) are frequently members of the minimal quorum}
\end{figure}

Voters on Compound and Uniswap can be partitioned into a few categories: investment firms (likely have investments here, try to check this), other DeFi-apps and companies, founders of other DeFi-apps and companies, founders and substantial contributors to these protocols, and university blockchain clubs. These categories identify the voters that have both an interest in these protocols and the necessary voting weight to be amongst the most powerful voters. Certainly the founders of Compound care about the governance of the platform, as do the leaders of the Uniswap Grants Program with Uniswap, and investment firms with financial stakes in these platforms. Although it is not quite as obvious, it makes sense that other participants in the DeFi ecosystem also care about the current well-being and future of these protocols, since they care about the ecosystem as a whole.

Additionally, there are entities that have been amongst the most powerful voters on Uniswap and Compound including a16z, Dharma Labs, Gauntlet, Robert Leshner, Blockchain at Berkeley, and Blockchain at UCLA. No discussion yet of the significance of these actors that are participating in multiple governance systems.

Some of the most active voters include

\begin{itemize}
\item \textbf{a16z}
Andreessen Horowitz (a16z) is a venture capital firm that invests in companies across bio and healthcare, consumer, crypto, enterprise, fintech, games, and other industries\footnote{\url{https://a16z.com/about/}}. A16z started investing in crypto in 2013, and has since raised \$7.6B in crypto/web3 funds \cite{a16zfund4}. The firm has led Series A investments for Uniswap \cite{uniswapseriesA} and Compound \cite{CompoundSeriesA}, and invested in Lido \cite{a16zLido}. In exchange for its investment, the firm received governance tokens, which it has used to vote on proposals itself and to delegate votes to other community members. a16z does delegate some portion of its votes, which has led to controversy in the Uniswap community as other community members have claimed a16z delegates decided entire votes and sought more transparency from a16z about their delegation strategy \footnote{\url{https://gov.uniswap.org/t/demand-for-transparency-from-defi-education-fund/13299}}.

Almost immediately following this complaint, a16z released their criteria for assessing delegates and their current delegate network. As of last year, a16z has delegated over half of its voting power on Compound and Uniswap. They have published their full assessment criteria \cite{delegaterubric}, but generally focus on the following areas: commitment to the protocol, subject matter expertise, history of engagement, independence from a16z, etc \cite{tokendelegate}. They also claim to be committed to providing delegates with enough voting weight to be able to make their own proposals and maintain their delegation for a minimum of 6 months even if the delegates vote against a16z's beliefs in every proposal during this period. On both Uniswap and Compound, a16z's current delegate network includes approximately 50\% of delegations to University Blockchain Clubs (e.g. Stanford, Penn, MIT, Berkeley), 26\% of delegations to startups (e.g. Dharma, Gauntlet, Argent), and approximately 20\% of delegations to non-profit organizations (e.g. Kiva, Mercy Corps).

Even given their delegated stake, a16z remains one of the most powerful voters on both Compound and Uniswap.

\item \textbf{Polychain capital}

Polychain capital is an investment firm focused on the blockchain space.  Although they have not invested Aave, Compound, Lido or Uniswap, they are active in the governance.

\item \textbf{Bain capital ventures}

Bain Capital Ventures (BCV) launched its \$560M Crypto Fund in March 2022 \cite{BCVfund}, and announced their intention of actively participating in DAO governance.
Bain participated in Compound's Series A \cite{CompoundSeriesA}.  Bain has not invested in the Aave, Lido or Uniswap, which may explain their lack of voting activity on those platforms.

\item \textbf{Gauntlet}
Gauntlet is a financial modeling and simulation platform for blockchains\footnote{\url{https://gauntlet.network/about/}}. Gauntlet applies agent-based simulation to measure protocol and smart contract behavior in order to help build secure and successful crypto systems. The platform's techniques can answer questions such as, ``Are people incentivized to participate in the system?,'' or ``What is the cost of a particular type of attack?'' \cite{introducinggauntlet} Two of Gauntlet's significant clients are Compound and Aave \cite{gauntletUnicorn}. Additionally, Gauntlet participates in governance as a delegate for token holders (I do not know if Gauntlet has its own token supply in a governance system).

\item \textbf{Dharma}
Dharma Labs built systems to convert fiat to crypto, including its Ethereum wallet, before it was bought by NFT marketplace OpenSea in January 2022 \cite{dharmaOpensea}. Dharma enabled users to directly deposit fiat into protocols such as Compound or Aave and exchange any asset in Uniswap. Before being acquired by OpenSea, Dharma Labs was active in Compound \cite{DharmaCompGov} and Uniswap governance \cite{DharmaUniGov,DharmaUniGov2}, advocating for retroactive airdrop of governance tokens to users who used these DeFi apps indirectly through Dharma's proxy smart contract (their efforts were not successful) \cite{dharmaRetroactive}.

\item \textbf{Robert Leshner}
Robert Leshner founded the DeFi lending platform Compound in 2017.  Although Robert Leshner still maintains a large number of COMP tokens\footnote{\url{https://compound.finance/governance/leaderboard}}, he is also extremely active in Uniswap governance.

\item \textbf{University Blockchain Clubs}
There are several university blockchain clubs that participate in on-chain governance as delegates. FranklinDAO (Penn), Blockchain at Berkeley, Blockchain at UCLA, Blockchain at Michigan, Stanford Blockchain Club, and Harvard Law BFI have all been present in minimal quorums.

\item \textbf{Monetsupply}
Monetsupply.eth is a ``delegate, risk analyst, and governance contributor for decentralized protocols.''\footnote{\url{https://www.monetsupply.xyz/home}} He participates in governance for 9 DeFi protocols, including Compound, Uniswap, MakerDAO, and Aave, and is a member of the MakerDAO risk core unit. Additionally, he holds investments in cryptocurrencies and governance tokens, including ETH, DAI, and MKR.

\item \textbf{Pantera Capital}
Pantera Capital is a hedge and venture fund that invests in various cryptocurrencies (including the first Bitcoin investment fund in 2013) and early-stage companies in the blockchain ecosystem. While the fund does not invest in any of the platforms in this case study, it has been in minimal quorums in Compound and Aave governance, which can be explained by their overall investments in the ecosystem as a whole.

\end{itemize}

\begin{figure}
    \centering
    \includegraphics{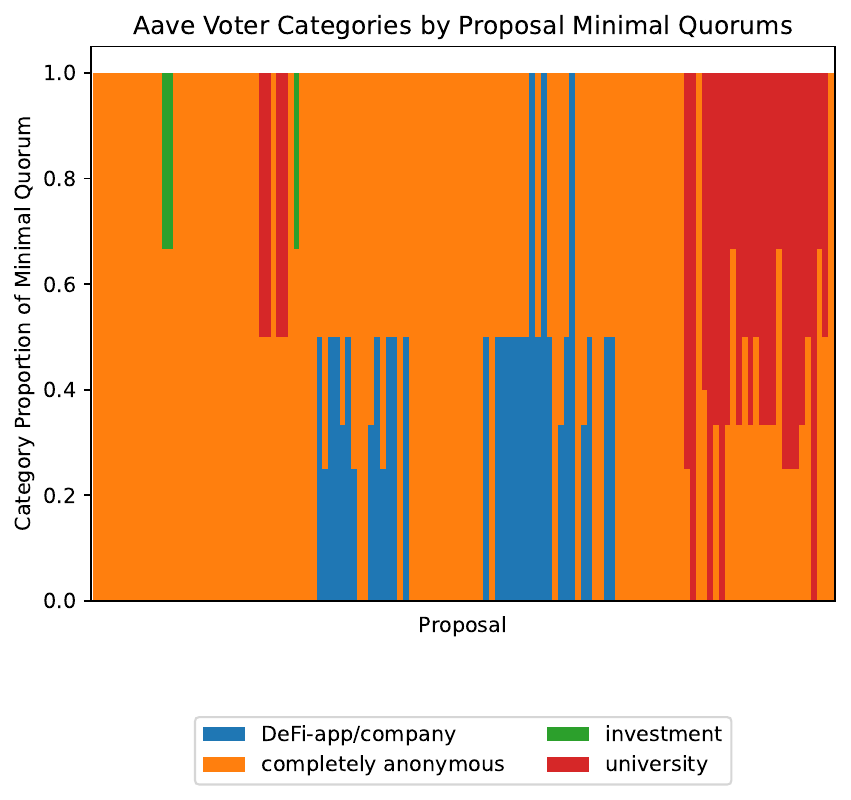}
    \caption{Participants in the Aave Minimal Quorums.  Although many of Aave's voters are anonymous, blockchain clubs have come to be dominant decision-makers on the Aave platform.}
    \label{fig:aave_categories}
\end{figure}

\begin{figure}
    \centering
    \includegraphics{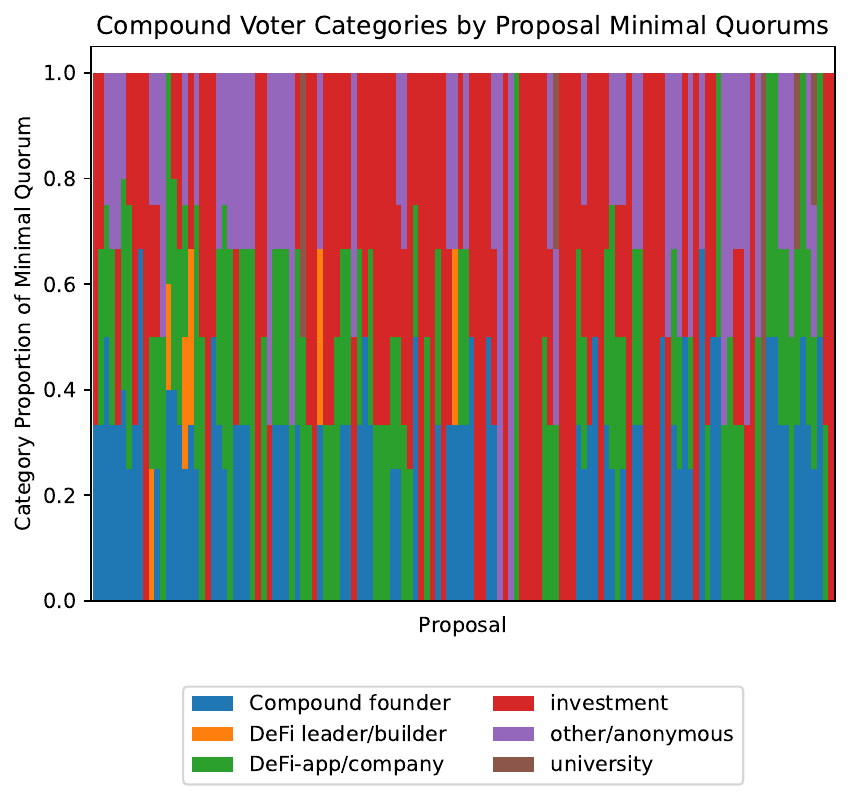}
    \caption{Participants in the Compound Minimal Quorums}
    \label{fig:comp_categories}
\end{figure}

\begin{figure}
    \centering
    \includegraphics{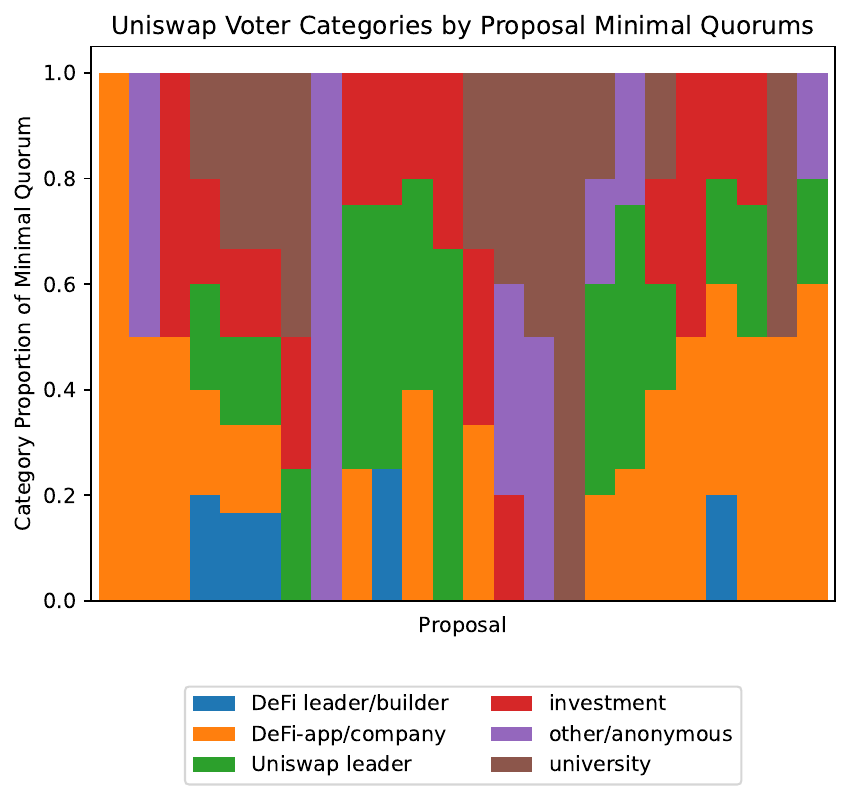}
    \caption{Participants in the Uniswap Minimal Quorums}
    \label{fig:uni_categories}
\end{figure}

Figures~\ref{fig:aave_categories}, \ref{fig:comp_categories} and \ref{fig:uni_categories} show the breakdown ``minimal quorums'' in Aave, Compound and Uniswap.  We do not present havbreakdowns for Lido because the majority of Lido DAO voters are anonymous, i.e., they are not registered with ENS, and unknown and their addresses are not labelled by sybil.org or Compound.

\bibliographystyle{alpha}
\bibliography{main}

\newpage
\appendix
\begin{center}
{\huge Appendix}
\end{center}
\section{Methodology}

In this work we collect on-chain data from four DeFi DAOs, Aave, Compound, Lido and Uniswap.
We collected vote data directly from the blockchain.
In particular, we ran our own Ethereum node, using the Erigon execution client and the Lighthouse consensus client.
This allowed us to pull raw data directly from the blockchain using Python's Web3 library.

Uniswap's governance contracts are a fork of Compounds, so the two systems are almost identical and the data collected 
from the two platforms is in similar formats.

Aave and Lido use different voting contracts.  Aave developed their own, custom governance contracts while Lido used the 
Aragon governance platform.

All of our tooling to collect and analyze this data is available \url{https://github.com/bhemen/dao_governance}.
\section{The Uniswap BNB chain vote}
\label{app:BNB}

In February 2023, Uniswap voted whether to deploy a version of Uniswap on the BNB chain\footnote{\url{https://www.tally.xyz/gov/uniswap/proposal/31}}.  The BNB chain uses the Ethereum Virtual Machine (EVM), so the Uniswap contracts could be deployed to BNB without modification.  To avoid fragmentation, however, the governance contracts (and UNI tokens) would not be recreated on the BNB chain.  Instead, all the governance decisions would remain on the Ethereum blockchain, and the results of those decisions would be sent to the Uniswap contracts on the BNB chain using a message-passing protocol.

This means that if Uniswap were to deploy on the BNB chain, they would have to a choose a message-passing protocol to relay messages between the governance contracts (which remain on Ethereum) and the newly deployed exchange contracts on BNB chain.
But which bridge would Uniswap use to pass governance decisions from Ethereum to BNB?  The bridge operator would \emph{not} stand to gain much in fee revenue for providing this service, however, being picked by Uniswap would be a huge stamp of approval for the protocol\footnote{\url{https://youtu.be/LHDeX7wYUts?t=3418}}.  Based on a preliminary ``temperature check,'' the proposal specified that Uniswap would use the Wormhole protocol to pass governance decisions (made on Ethereum) to the Uniswap contracts on BNB.  The Wormhole protocol\footnote{Wormhole is the message-passing layer powering the Portal Bridge https://wormholecrypto.medium.com/wormhole-protocol-vs-portal-bridge-whats-the-difference-a9ad98696e64
} has been backed by the VC firm Jump Capital\footnote{\url{https://decrypt.co/92709/jump-crypto-wormhole-defi}}.  One of the other message-passing layers considered was LayerZero, a protocol backed by the VC firm a16z\footnote{\url{https://a16z.com/2022/03/30/investing-in-layerzero/}}.  a16z is a large UNI token holder, and voted against this proposal, instead favoring launching Uniswap on BNB using the LayerZero protocol.  Although a16z voted against the proposal, several of its delegates (mostly university clubs) voted for the proposal, and the proposal ultimately passed.

The drama has called into question whether VCs like a16z are actually in control of these purportedly decentralized projects\footnote{\url{https://decrypt.co/120653/uniswap-controlled-a16z-crypto-twitter-split-over-vc-firms-governance-move}}.

\end{document}